\documentclass{aastex}
\usepackage{emulateapj5,apjfonts}
\shorttitle{ORBITAL PERIOD OF 4U 1543--624}
\shortauthors{WANG \& CHAKRABARTY}

\begin{document}
\bibliographystyle{apj_noskip}

\title{The Orbital Period of the Ultracompact Low-Mass X-Ray
Binary 4U 1543--624} 

\author{Zhongxiang Wang and Deepto~Chakrabarty\altaffilmark{1}}
\affil{\footnotesize Department of Physics and Center for Space Research, 
   Massachusetts Institute of Technology, \\ Cambridge, MA 02139}
\email{wangzx@space.mit.edu, deepto@space.mit.edu}

\altaffiltext{1}{Alfred P. Sloan Research Fellow.}

\begin{abstract}
We report the discovery of the orbital period of the ultracompact 
low-mass X-ray binary (LMXB) 4U~1543$-$624 using time-resolved 
optical photometry taken 
with the 6.5-m Clay (Magellan II) telescope in Chile. The light curve 
in the Sloan $r'$ band clearly shows a periodic, sinusoidal
modulation at 18.2$\pm0.1$ min with a fractional semiamplitude
of 8\%, which we identify as the binary period. This is the second
shortest orbital period among all the known LMXBs, and it verifies
the earlier suggestion of 4U~1543$-$624 as an ultracompact binary
based on X-ray spectroscopic properties. The sinusoidal shape of
the optical modulation 
suggests that it arises from X-ray heating of the mass donor
in a relatively low-inclination binary, although it could also be a superhump
oscillation in which case the orbital period is slightly shorter.  
If the donor is a C-O white dwarf as previously suggested, its likely
mass and radius are around 0.03 $M_{\sun}$ and $0.03\ R_{\sun}$, 
respectively. For conservative mass transfer onto a 1.4 $M_{\sun}$ 
neutron star and driven by gravitational radiation,
this implies an X-ray luminosity of 6.5$\times 10^{36}$ erg s$^{-1}$
and a source distance of $\approx$7 kpc.
We also discuss optical photometry of another LMXB, the candidate
ultracompact binary 4U~1822$-$000. We detected significant optical
variability on a time scale of about 90~min, but it is not yet clear
whether this was due to a periodic modulation.

\end{abstract}


\section{INTRODUCTION}

Low-mass X-ray binaries (LMXBs) containing ordinary, hydrogen-rich
mass donors have a minimum orbital period around 80~min (Paczy\'{n}ski \&
Sienkiewicz 1981; Rappaport, Joss, \& Webbink 1982).  However, systems
with hydrogen-poor or degenerate donors can evolve to extraordinarily
small binary separations, with orbital periods as short as a few minutes
(Nelson, Rappaport, \& Joss 1986).  These so-called ultracompact
binaries include three X-ray bursters (two in globular clusters), a
classical X-ray pulsar, and three millisecond X-ray pulsars, spanning
a range of orbital periods from 11 to 50 minutes.  Besides these
accreting neutron stars, there is also a related class among the
accreting white dwarfs, the AM CVn binaries (see Warner 1995).
Together, these systems represent extreme and exotic endpoints in
binary and stellar evolution (see, e.g., Podsiadlowski, Rappaport, \&
Pfahl 2002).  In all cases, the donor stars in these systems must have
extremely low mass and be either hydrogen-depleted or degenerate
(Nelson, Rappaport, \& Joss 1986; Yungelson, Nelemans, \& van den
Heuvel 2002).
                                                                                
Although these systems had initially been assumed to be relatively
rare, the number known has doubled in the past few years and theoretical
studies indicate that they could be more common than previously realized
(see, e.g., Belczynski \& Taam~2003).  In addition,
recent X-ray spectroscopic work has identified several more candidate
ultracompact binaries on the basis of comparison to the known
ultracompact LMXB 4U 1626$-$67, with low-mass, Ne-enriched C-O white
dwarfs suggested as possible donors (Juett, Psaltis, \& Chakrabarty 2001;
Juett \& Chakrabarty 2003).  Candidates may also be identified through
an unusually low optical--to--X-ray flux ratio 
(van Paradijs \& McClintock~1994; Deutsch, Margon, \& Anderson~2000).
In an effort to verify these proposed candidates, we have recently
undertaken a systematic optical survey aiming to detect orbital flux
modulations through time-resolved photometry.  We report on the first
results of our survey in this paper with observations of the LMXBs
4U~1543$-$624 and 4U~1822$-$000.
                                                                                
These two sources were discovered by the {\em Uhuru} mission over thirty
years ago (Giacconi et al. 1972).  Both are presumed to be accreting
neutron stars, although the absence of either X-ray bursts or pulsations
precludes a definitive conclusion.   The source 4U 1543$-$624
($l=322^\circ$, $b=-6^\circ$) has since been extensively observed by a
series of X-ray missions (Singh, Apparao, \& Kraft 1994; Christian \&
Swank 1997; Asai et al. 2000; Schultz 2003; Farinelli et al. 2003;
Juett \& Chakrabarty 2003).  It was identified as a candidate
ultracompact system by Juett et al. (2001), but observations with the
{\em Chandra X-Ray Observatory} and {\em XMM-Newton} found no evidence
for orbital modulation of the X-ray flux (Juett \& Chakrabarty 2003).
The $B\simeq 20$ optical counterpart of 4U~1543$-$624 was identified
by McClintock et al. (1978) based on its {\em SAS-3} position, which
has been subsequently verified by {\em Chandra} (Juett \& Chakrabarty
2003).  Optical  spectra show no lines of H or He and support the
suggestion of C-O white dwarf donor (Nelemans et al. 2004; Wang \&
Chakrabarty 2004).   The X-ray source 4U~1822$-$000 ($l=30\arcdeg$,
$b=+6\arcdeg$) has been less well
studied.  Its $V=22$ optical counterpart was identified by Chevalier
\& Ilovaisky (1985).   {\em Chandra} observations verify that the
X-ray source is coincident with the optical position and show no
evidence for orbital modulation of the X-ray flux (Juett \&
Chakrabarty 2004).  We identified the system as a candidate
ultracompact system on the basis of its optical/X-ray flux ratio.
In this paper, we report the detection of an 18.2-minute periodicity
in the optical flux from 4U~1543$-$624 and the presence of significant
variability in the optical flux from 4U~1822$-$000.

\section{OBSERVATIONS AND DATA REDUCTION}    

Our optical photometric observations were made on 2003 August 2--4 using  
the 6.5-meter Clay/Magellan~II telescope at Las Campanas
Observatory in Chile. The detector was
the Raymond and Beverly Sackler Magellan Instant Camera (MagIC),
a 2048$\times$2048 pixel CCD
camera providing a 0\farcs069 pixel$^{-1}$ plate 
scale and a 142\arcsec\
field of view at the f/11 focus of the telescope. 
A Sloan $r'$ filter (Fukugita et al.~1996) was used for our observations.
The total observation time spanned approximately 
100 min on August 2 for
4U 1822$-$000 with 99 CCD image frames of the targeted field taken, 
and 140 min on August 3 for 4U 1543$-$624 with 137 frames taken. 
The exposure time of each individual frame was 30 seconds for both
targets. Since the readout time of MagIC is 20 seconds, 
we obtained approximately one image per minute over the course of our 
observations. The telescope position was dithered 5\arcsec\ once every 20 
minutes during
the observations. The conditions during our observations on August 2 
were excellent, with 
the 0\farcs6 seeing. On August 3, the conditions were
windy, with the seeing varying in a range of 0\farcs6--1\farcs0. 
In addition to our science targets, the Sloan photometric
standard star G~93-48 (Smith et al.~2002)
was observed on August 3 for flux calibration of 4U 1543$-$624.
A few images of both science targets were obtained on August 4,
allowing us to extend our photometric calibration to 4U 1822$-$000 as well
with several non-variable stars in the field of 4U 1543$-$624 as standard stars.
In Figure 1, we show finder images of our two fields.
\begin{figure*}
\includegraphics[scale=0.6]{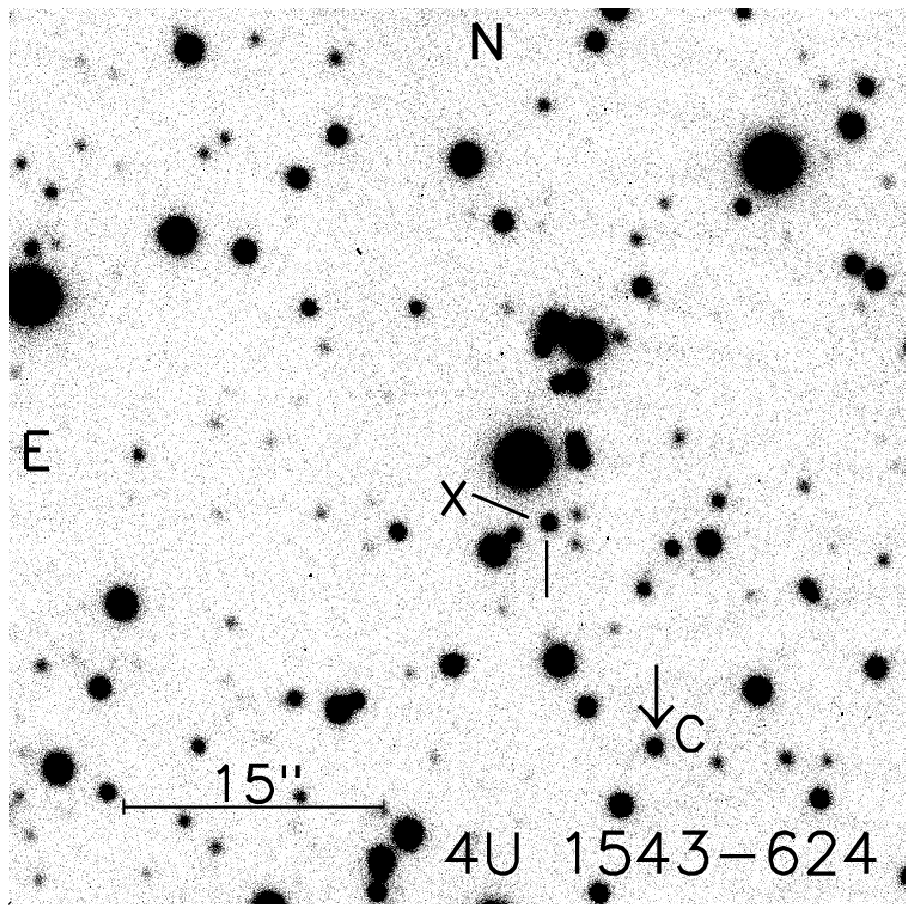}
\includegraphics[scale=0.6]{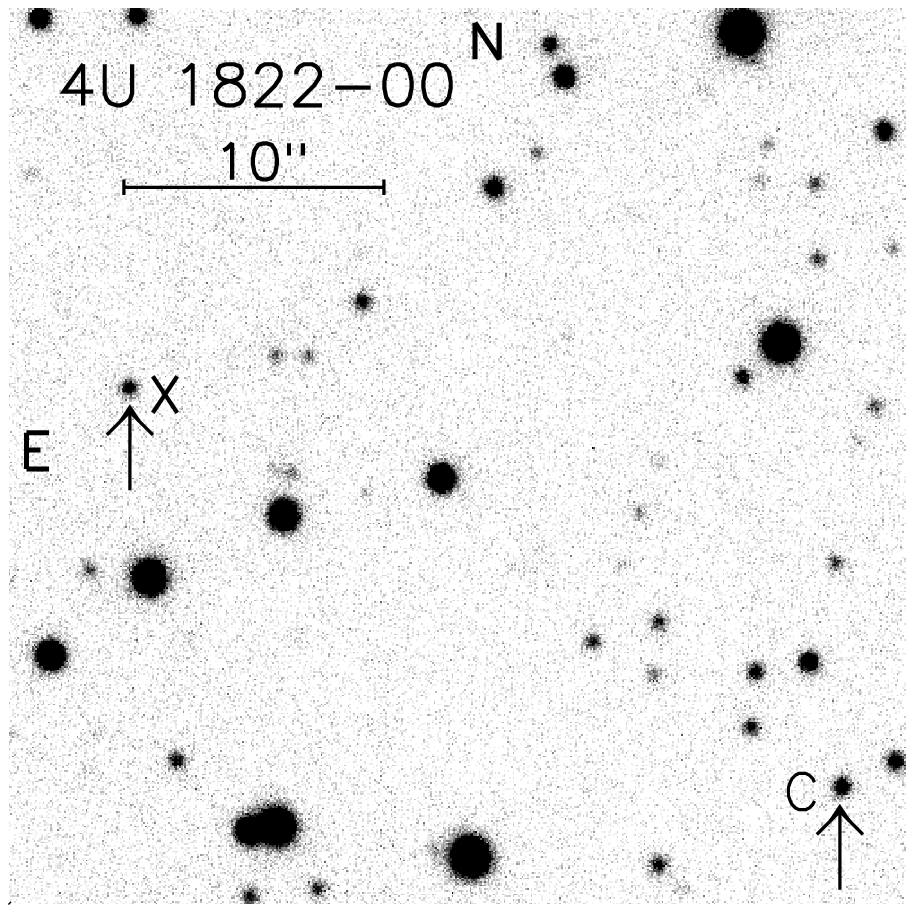}
\caption{Sloan $r'$-band images of the fields containing 4U~1543$-$624
and 4U~1822$-$000 with their optical counterparts indicated 
(object {\em X}; $r'\approx 20.4$ and 21.6, respectively). Also indicated
are check stars (object {\em C}) for comparison of light curves.}
\end{figure*}

We used the {\tt IRAF} analysis package for our initial data reduction,
including bias subtraction and flat fielding. 
We then used {\tt DOPHOT} (Schechter, Mateo, \& Saha~1993), 
a point-spread function fitting photometry program, 
to measure brightnesses of 
our target and other in-field stars. In order to eliminate any
systematic offsets, we used differential photometry 
for our timing analysis;
the brightnesses of our targets were calculated relative to an ensemble of 
bright, non-variable stars in the images. 

\section{RESULTS}

The light curve for 4U~1543$-$624 is shown in the top panel of Figure 2. 
The average 
magnitude was $r'= 20.42\pm 0.03$. However,
a periodic modulation with a semiamplitude of around 
\begin{center}
\includegraphics[scale=0.5]{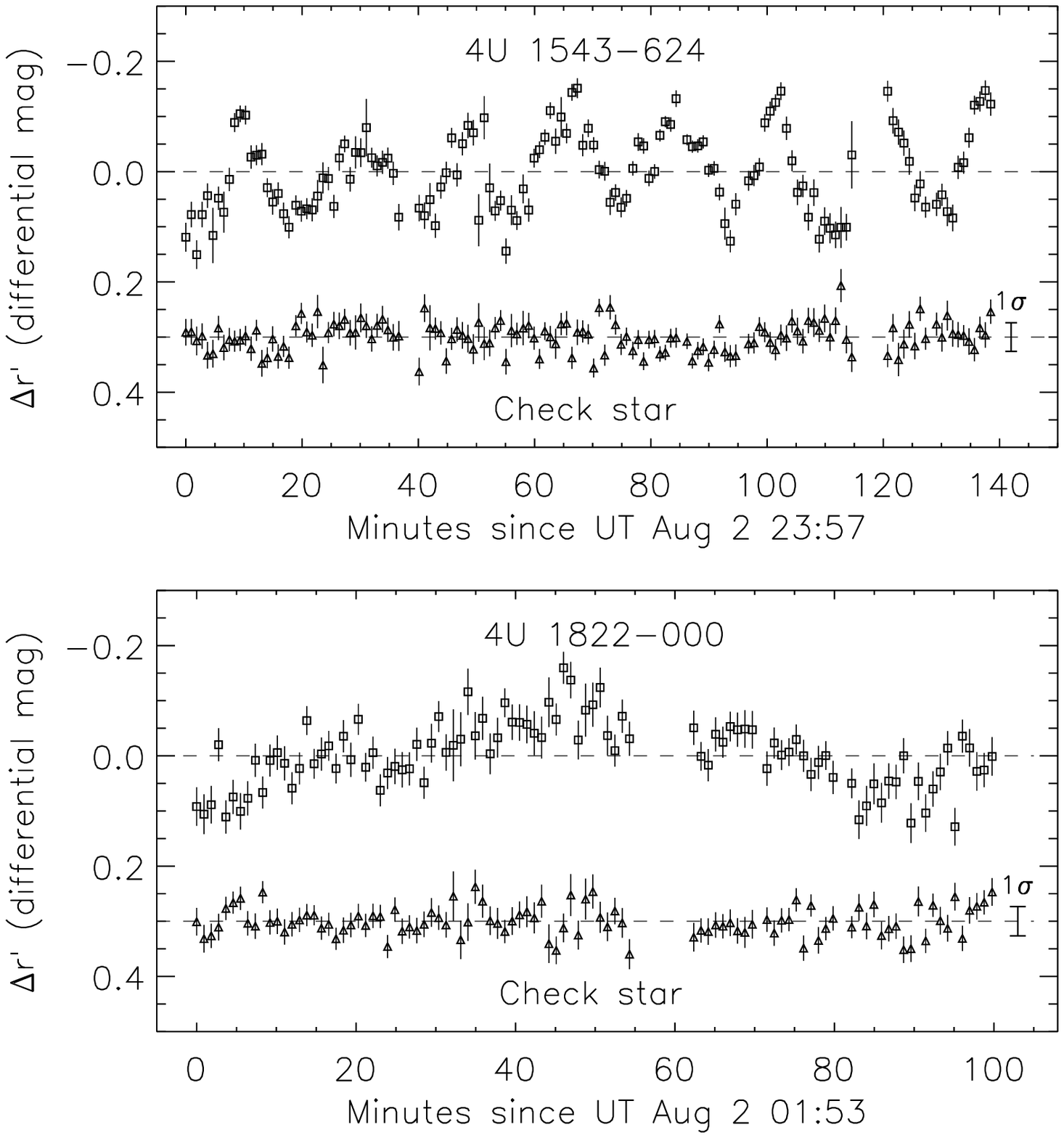}
\figcaption{ Light curves of 4U~1543$-$624 and 4U~1822$-$000 in the $r'$ band.
The light curves for nearby comparison stars with comparable brightness
are also shown. A periodic modulation is clearly visible in the light
curve of 4U~1543$-$624. The brightness of 4U~1822$-$000
varies significantly over the observation.
 The average magnitudes were $r'=20.42\pm0.03$ for
4U~1543$-$624 and $r'=21.58\pm0.08$ for 4U~1822$-$000.}
\end{center}
0.1 magnitudes
is also clearly visible.  For comparison, the light curve of a check 
star of similar brightness in the same field is also shown (see Figure~1). 
The standard deviation
of the brightness of this check star is only 0.026 magnitudes, which shows 
that the light curve variation of our target is highly significant.
We made an initial estimate of the modulation period by interpolating
the data into an evenly-sampled time series and using a Fourier
spectral analysis, which indicated that the modulation was periodic and
essentially sinusoidal. We refined our measurement using an epoch-folding
search technique on the uninterpolated data (Leahy et al.~1983) to 
determine the period of
18.2$\pm 0.1$~minutes. A plot of the data folded on this period
is shown in Figure~3, along with the best-fit sinusoid with a semiamplitude
of 0.081$\pm 0.002$ magnitudes (the reduced $\chi^2$ is 1.8 for 133 d.o.f.). 
The topocentric time of phase zero
(maximum brightness) was UT 2003 August 3 00:44:20, corresponding
to August 3 00:48:32 (TDB) at the solar system barycenter, with an uncertainty
of 66 seconds.

In the bottom panel of Figure~2, we show the light curve of 4U 1822$-$000. 
The average magnitude was $r'=21.58\pm0.08$.
However, the target brightness is clearly varying systematically
by a few tenths of a magnitude over the 100~min observation.
These changes are quite significant compared to the 0.027 magnitude
scatter observed in nearby check star. The variability could
possibly be due to a $\sim$90~min periodicity, but our data
span is insufficient to determine this.
\begin{center}
\includegraphics[scale=0.65]{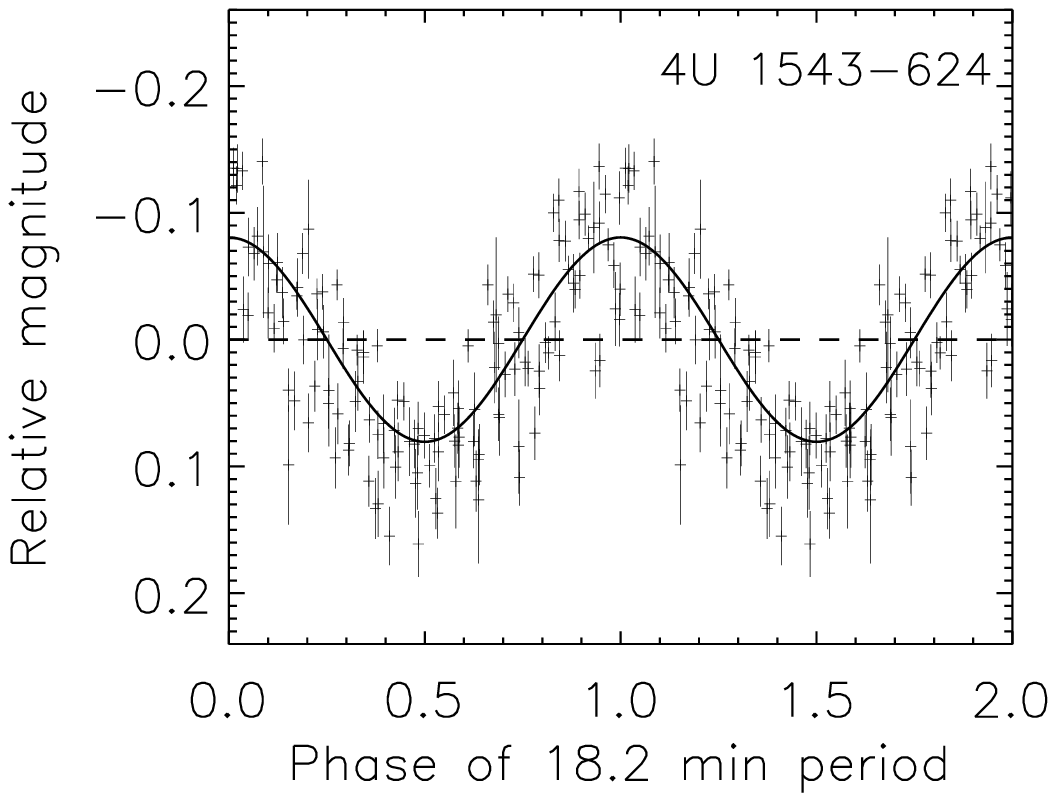}
\figcaption{
The $r'$ band photometric data for 4U 1543$-$624 folded on the 18.2 min period.
Phase zero is chosen
to be of maximum brightness. Two cycles are displayed for clarity.
The best fitting sinusoid is plotted as the solid curve.}
\end{center}

\section{Discussion}

We have found an 18.2~min periodicity in the optical light curve of
4U~1543$-$624.  Since the modulation appears to be coherent, this is
almost certainly the orbital period of the binary. 
Although we cautiously note that the periodic modulation
could arise from the accretion disk (see discussion below), we
have verified the ultracompact nature of 4U~1543$-$624.
This is the second shortest
orbital period for an LMXB, after the 11-minute binary 4U~1820$-$30
(Stella, Priedhorsky, \& White 1987).  We note that this discovery
bolsters the earlier suggestion of
4U~0614+091 and 2S~0918$-$549 as ultracompact binaries on the basis of
X-ray and optical evidence (Juett et al. 2001; Juett \& Chakrabarty
2003; Nelemans et al. 2004; Wang \& Chakrabarty 2004).

Sinusoidal
modulation of the optical light curve for LMXBs generally arises from
X-ray heating of the companion star by the central X-ray source in a
relatively low-inclination binary (no X-ray eclipse or dip; 
see van Paradijs \& McClintock 1995), with the visible area of the heated 
face varying as a function
of orbital phase, and the superior conjunction of the companion star
corresponding to the observed brightness maximum of the light curve.
For sufficiently low binary inclinations (generally $i < 60\arcdeg$;
Frank, King, \& Raine 1993), blockage by the accretion
disk does not occur, resulting in a sinusoidal profile.  A possible
alternative explanation is that the 18.2~min variation is a superhump
oscillation (see Warner 1995 for a review).  Superhumps are observed
as optical photometric modulations in cataclysmic 
variables and LMXBs
at periods a few percent longer than the binary period.  These
oscillations, which only occur in binaries with extreme mass ratios
(like ultracompact binaries), are understood to arise because of an
orbital resonance condition 
that leads to a precessing, eccentric
accretion disk (Whitehurst 1988; Whitehurst \& King 1991; Lubow 1991).
Without an independent determination of the binary period from, e.g.,
X-ray variability or Doppler line measurements, we cannot definitively
distinguish between these possibilities.  In either case, however, 
4U~1543$-$624 certainly has an orbital period around 18~min.  
\begin{center}
\includegraphics[scale=0.65]{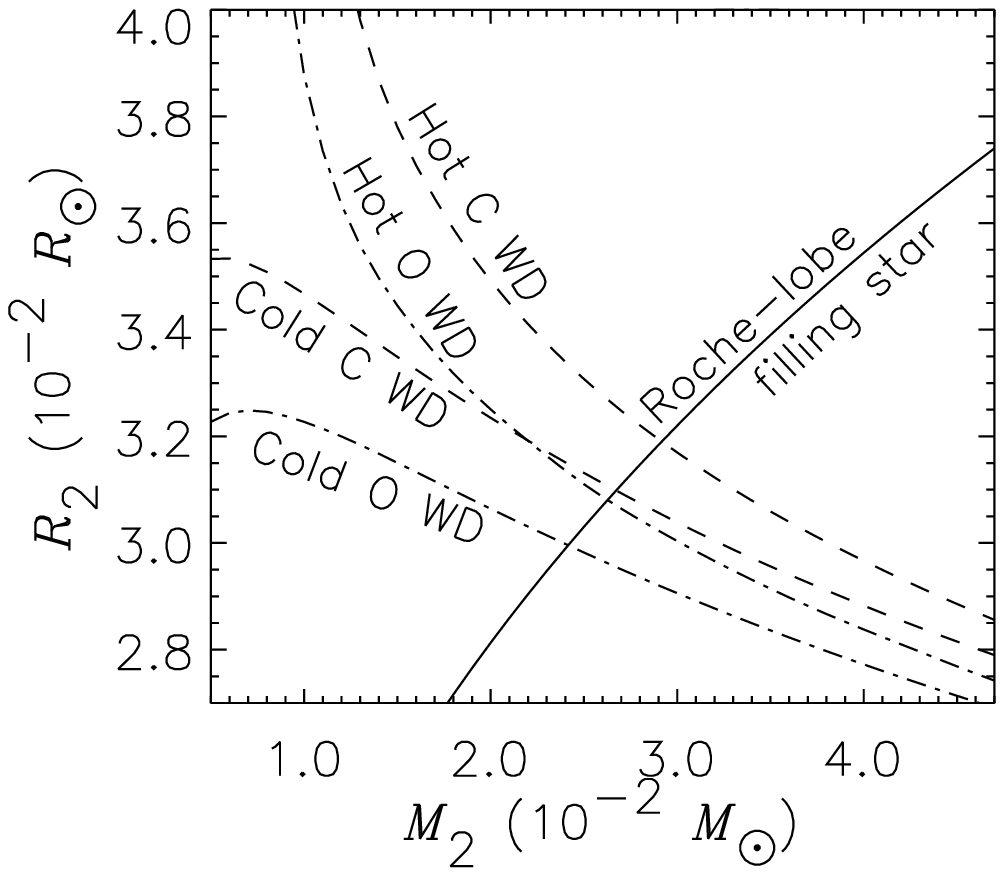}
\figcaption{Mass-radius constraints for the companion star in
4U~1543$-$624. The solid curve is the $M$-$R$ relation for a
Roche-lobe--filling donor in an 18.2~min binary.  Also shown are the
model curves for low-mass carbon (dashed lines) and oxygen (dot-dashed
lines) white dwarfs for both cold (10$^4$ K) and hot (3$\times 10^6$ K)
core temperatures, taken
from Deloye \& Bildsten (2003).  The donor must have a mass in the
0.025--0.03~$M_\odot$ range with a radius around 0.03~$R_\odot$.}
\end{center}

As an ultracompact binary, 4U 1543$-$624 must contain a
hydrogen-depleted or degenerate donor resulting from the evolution of
either an evolved main-sequence star+neutron star binary or a
white dwarf+neutron star binary (see Nelson \& Rappaport~2003 and
references therein for a recent discussion of evolutionary scenarios
for ultracompact binaries).  Indeed, based on both X-ray (Juett et
al. 2001; Juett \& Chakrabarty 2003) and optical measurements
(Nelemans et al. 2004; Wang \& Chakrabarty 2004), it has been
suggested that the donor is a low-mass C-O white dwarf.  If so, we can
use our orbital period measurement to estimate the mass and radius of
the donor.  Since the mean density of a Roche-lobe--filling companion
is determined by the binary period, our 18~min period defines a
mass-radius relation for the companion, shown as the solid curve
in Figure~4.  (In the absence of measured mass function for the
binary, the allowed curve extends down to $M_2=0$.)  
In comparing this
to stellar models, we note that recent studies of three millisecond
pulsars in ultracompact LMXBs have shown that the extremely low-mass
white dwarf donors in such systems may be thermally bloated compared
to cold stars, affecting their $M$-$R$ relation (Bildsten 2000; Deloye
\& Bildsten 2003).  For comparison, Figure~4 also shows both cold and
hot solutions for pure C and O white dwarfs from the models of Deloye
\& Bildsten (2003).  For a Roche--lobe-filling donor, a mass in the
0.025--0.03~$M_{\odot}$ range and a radius of 0.030--0.032~$R_{\odot}$ is
indicated.  Since mass transfer in an ultracompact binary is driven by
gravitational radiation, this mass estimate implies a mass transfer
rate of
\[
    \dot M \approx 5.5\times 10^{-10} M_\odot \mbox{\rm \ yr$^{-1}$}
     \left(\frac{M_1}{1.4\ M_\odot}\right)^{2/3}
     \left(\frac{M_2}{0.03\ M_\odot}\right)^2
\]
\[
   \times  \left(\frac{P_{\rm orb}}{18.2 \mbox{\rm\ min}}\right)^{-8/3} ,
\]
where $M_1$ is the mass of the compact primary, $M_2$ is the mass of
the white dwarf donor, and $P_{\rm orb}$ is the binary period.  Assuming
conservative mass transfer onto a 1.4 $M_{\sun}$ neutron star,
the measured unabsorbed 0.5--10~keV X-ray flux of $\simeq1\times
10^{-9}$ erg~cm$^{-2}$~s$^{-1}$ (Juett \& Chakrabarty 2003)
suggests a source distance of $\simeq$7~kpc.

The $\sim$90~min optical variability of 4U~1822$-$000 
is clearly significant. However, given our limited data span,
it is unclear whether it is periodic, quasiperiodic, or 
stochastic.  Strong $\sim$15~min optical/UV quasiperiodic oscillations
were previously detected in the 42~min ultracompact LMXB 4U~1626$-$67
(Chakrabarty et al. 2001), showing that photometric variability in
ultracompact binaries need not only occur near the orbital period.
Again, only an observation long enough to contain many modulation
cycles can distinguish between a periodic and a quasiperiodic
oscillation (or stochastic variability) and allow a secure measurement
of its time scale.

\acknowledgements{We thank Paul Schechter for obtaining the data for
us and suggesting we use the {\tt DOPHOT} photometry program, Adrienne
Juett for useful discussion, Jennifer
Sokoloski for advice on timing analysis, and Christopher Deloye
and Lars Bildsten for sharing their white dwarf models with us.
This work was supported in part by the Alfred P. Sloan Foundation.}


\clearpage

\end{document}